\documentclass[conference]{IEEEtran}
\IEEEoverridecommandlockouts
\usepackage{cite}
\usepackage{amsmath,amssymb,amsfonts}
\usepackage{algorithmic}
\usepackage{graphicx}
\usepackage{textcomp}
\usepackage{xcolor}
\def\BibTeX{{\rm B\kern-.05em{\sc i\kern-.025em b}\kern-.08em
    T\kern-.1667em\lower.7ex\hbox{E}\kern-.125emX}}
\begin{document}

\title{Towards Advanced Speech Signal Processing: A Statistical Perspective on Convolution-Based Architectures and it's Applications
}

\author{\IEEEauthorblockN{\textbf{Kapu Nirmal Joshua}}
\IEEEauthorblockA{\textit{nirmalj21@iitk.ac.in}}
\textit{Department of Electrical Engineering, IIT Kanpur} \\

\and
\IEEEauthorblockN{\textbf{Raghav Karan}}
\IEEEauthorblockA{\textit{raghavk20@iitk.ac.in}} 
\textit{Department of Electrical Engineering, IIT Kanpur}\\
}

\maketitle

\begin{abstract}

This article surveys convolution-based models-convolutional neural networks (CNNs), Conformers, ResNets, and CRNNs-as speech signal processing models and provide their statistical backgrounds and speech recognition, speaker identification, emotion recognition, and speech enhancement applications. Through comparative training cost assessment, model size, accuracy and speed assessment, we compare the strengths and weaknesses of each model, identify potential errors and propose avenues for further research, emphasising the central role it plays in advancing applications of speech technologies.

\end{abstract}


\section{Introduction}

In this subsection, the principle of convolution and its application to the modification of a speech signal are explained.

\subsection{Mathematical Foundation for Convolution}

Most broadly, convolution is the mathematical process of compounding two functions to produce a third function, which reflects the influence of one effect on the other—how changing one alters the form of the other. Convolution is effective for analyzing linear time-invariant (LTI) systems and is applied in a wide range of engineering fields, such as speech processing \cite{b1}.

\begin{equation}
    (f \ast g)(t) = \int_{-\infty}^{\infty} f(\tau) \cdot g(t - \tau) \, d\tau
    \label{eq:continuous_convolution}
\end{equation}

Where $(f \ast g)(t)$ is the convolution of $f(\tau)$ and $g(t - \tau)$, where both functions overlap in all time.

The discrete convolution formula for sequences $\{f[k]\}$ and $\{g[k]\}$ is as follows:

\begin{equation}
    (f \ast g)[n] = \sum_{k=-\infty}^{\infty} f[k] \cdot g[n - k].
    \label{eq:discrete_convolution}
\end{equation}

Equation \eqref{eq:discrete_convolution} defines the discrete convolution, which sums products of $f[k]$ and shifted versions of $g[k]$ for all integers $k$.

If $x(t)$ is the input of the system and $h(t)$ is its impulse response, then $Y(t)$ has the following form:

\begin{equation}
    Y(t) = (x \ast h)(t) = \int_{-\infty}^{\infty} x(\tau) \cdot h(t - \tau) \, d\tau
    \label{eq:system_response}
\end{equation}

This equation represents the output signal as a function of the input signal modified by the system's impulse response.

There are several useful properties for signal processing analysis that convolutions possess. It is commutative, i.e., $f \ast g = g \ast f$; associative, i.e., $f \ast (g \ast h) = (f \ast g) \ast h$; and distributive over addition, i.e., $f \ast (g + h) = (f \ast g) + (f \ast h)$. These characteristics allow the reduction and analysis of complex systems.

\subsection{Introduction to Speech Signal Processing}

Speech signal processing involves analyzing speech signals and developing techniques to process them for effective communication between humans and machines. Speech signals are 1D, time-varying signals that are a manifestation of acoustic description of human language \cite{b1}. These signals are generally (a) non-stationary, (b) large dynamic range, and (c) rich spectral content, which can be challenging to analyze.

In speech signal processing, convolution plays an important role. When a speech signal $s(t)$ is transmitted or recorded in a communication channel, it is changed by the channel impulse response $h(t)$. Received signal $r(t)$ can be expressed as a convolution between speech signal and channel impulse response:

\begin{equation}
    R(t) = (s \ast h)(t) = \int_{-\infty}^{\infty} s(\tau) \cdot h(t - \tau) \, d\tau
    \label{eq:received_signal}
\end{equation}

This equation models the interaction between the speech signal and the channel, incorporating effects such as echo, reverberation, and attenuation.

If noise $n(t)$ is added, the received signal becomes:

\begin{equation}
    r(t) = s(t) \ast h(t) + n(t)
    \label{eq:received_signal_noise}
\end{equation}

The existence of noise makes it difficult to reconstruct the priori speech signal.

Convolution is also employed for feature extraction in the speech domain, such as calculating the frequency content of speech. The filtered speech signal can be convolved with a sequence of filters to realize time-frequency representations such as short-time Fourier transform, which are important for speech recognition and speaker labeling \cite{b1}, \cite{b13}. These methods derive articulatory and prosodic features of speech that are of paramount importance to speech decoding of spoken language.

Further, speech signal heterogeneity—that is, heterogeneity of the signals arising from speakers, accents, speech rate, and emotional states—introduces the challenge of complexity. Convolutional approaches must account for this heterogeneity. Yet, the issues, e.g., background noise, reverberation and real-world speech signal mixing, continue to be problems. Convolutional approaches, augmented with statistical signal processing, is an active research area for achieving robustness and performance in speech processing systems \cite{b9}, \cite{b19}. These methods are also the basis of developments in voice-based technology, automatic transcription and voice biometrics.

\subsection{Convolution-Based Architectures}

Recent breakthroughs in deep learning have enabled deep and convolutional based architectures which address a variety of problems in speech signal processing. The 4 functional architectures investigated are Convolutional Neural Network (CNN), Conformers, Convolutional Recurrent Neural Network (CRNN), and Residual Networks (ResNets).

\begin{figure}[!t]
    \centering
    \includegraphics[width=0.4\textwidth]{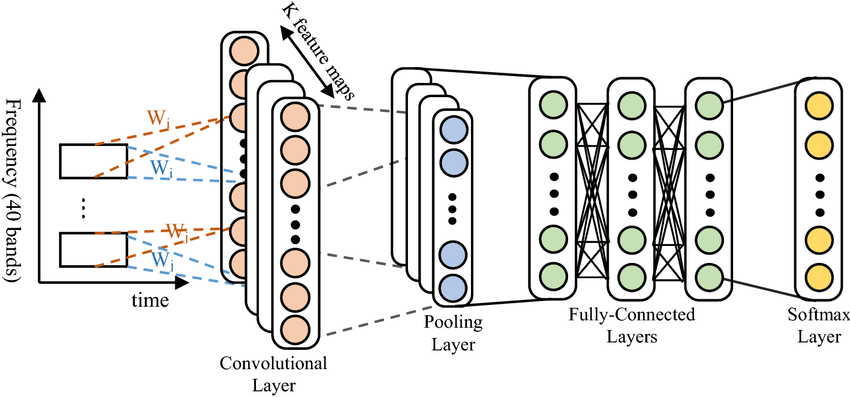}
    \caption{Convolution-Based Architectures}
    \label{fig:architectures}
\end{figure}

In these architectures, convolutional layers are used to successively extract features of speech signals. Convolutional neural networks (CNNs) apply convolutional layers for fast feature extraction, whereas Convolutional Complex Architectures (Conformers) apply convolution along with self-attention for the local and global relationships. CRNNs combine convolutional and recurrent layers to learn temporal dynamics, and ResNets exploit shortcut connections to build deeper architectures without performance overfitting. Collectively, these architectures increase the accuracy of speech signal processing, including in noisy and dynamic environments.

\section{Convolution Based Architectures}

\subsection{Convolutional Neural Networks (CNNs)}

Convolutional neural networks (CNNs) have almost universal relevance for the analysis of structured data, e.g., spectrograms, in vocal signal processing. In convolutional neural networks (CNNs) layers with trainable filters convolutional layers are employed to slide across the input data and learn local patterns. The one-dimensional convolution operation is written as follows with a temporal input sequence \( x(t) \) and a convolution kernel \( w(k) \):

\begin{equation}
y(t) = \sum_{k=-K}^{K} w(k) \cdot x(t - k)
\label{eq:1d_convolution}
\end{equation}

Output \( y(t) \) is defined where temporal dependencies are encoded, which are the basis for efficient speech signal processing \cite{b3}. CNNs make use of two-dimensional convolutions in time and frequency, which are as follows when applied to spectrograms:

\begin{equation}
y(i, j) = \sum_{m=-M}^{M} \sum_{n=-N}^{N} W(m, n) \cdot S(i - m, j - n),
\label{eq:2d_convolution}
\end{equation}

where \( S(i, j) \) is the spectrogram value at time \( i \) and frequency \( j \), and \( W(m, n) \) is a filter. This lets CNNs also learn the frequency features specific to speech recognition tasks \cite{b22}.

\begin{figure}[htbp]
\centering
\includegraphics[width=0.5\linewidth]{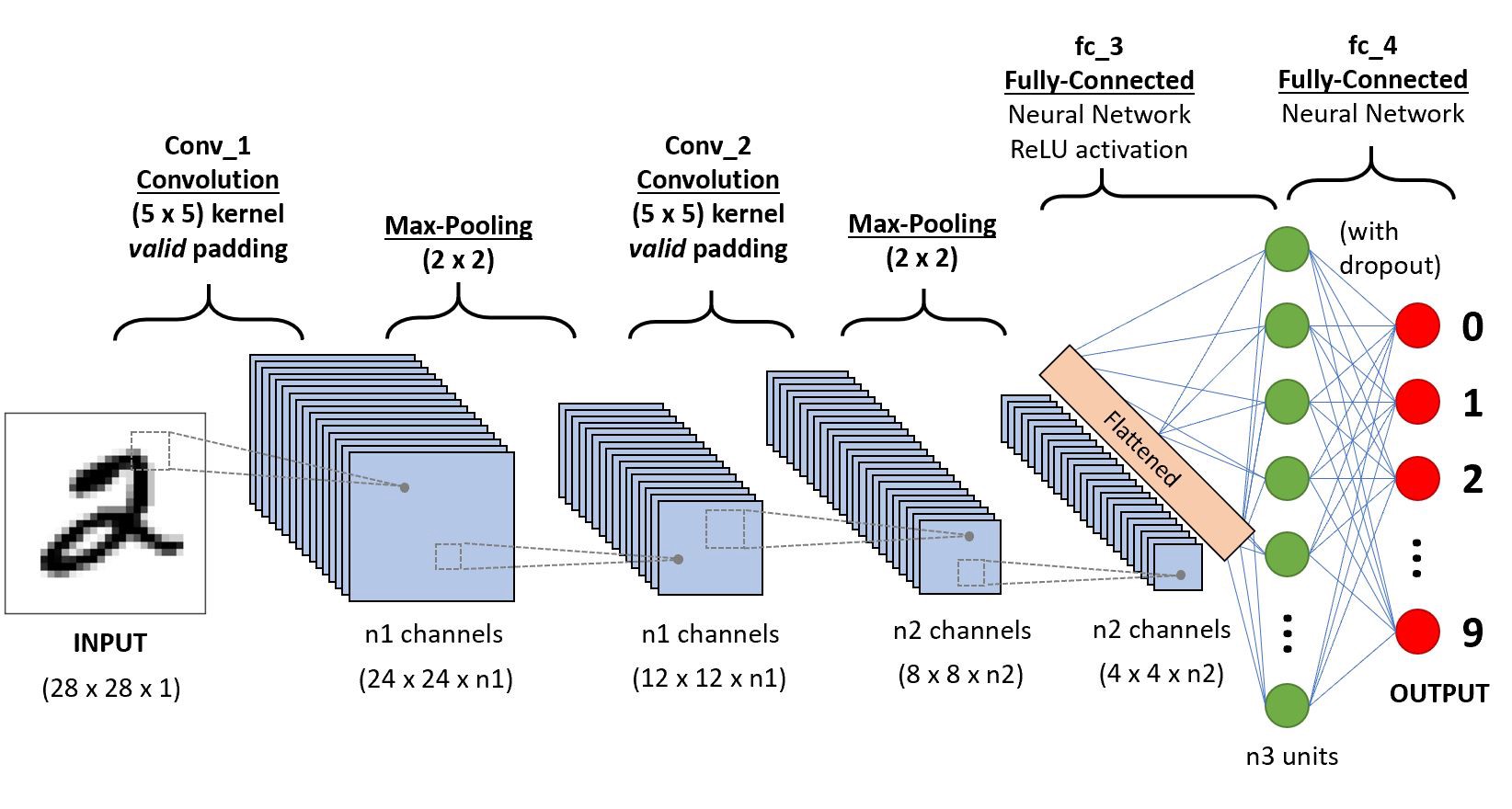}
\caption{Illustration of a CNN architecture in speech processing.}
\label{fig:cnn_architecture}
\end{figure}

Pooling layers, e.g., max pooling, downsample spatial resolution and retain global features (i.e., lead to higher translation invariance). Given a feature map \( y(i, j) \), max-pooling over a window size \( p \times q \) is defined as:

\begin{equation}
z(i, j) = \max_{0 \leq m < p} \max_{0 \leq n < q} y(i + m, j + n).
\label{eq:max_pooling}
\end{equation}

Batch normalization also stabilizes and speeds up training by normalizing layer outputs. For an activation \( a \) with mean \( \mu \) and variance \( \sigma^2 \), the normalized output is:

\begin{equation}
\hat{a} = \frac{a - \mu}{\sqrt{\sigma^2 + \epsilon}}, \quad y = \gamma \cdot \hat{a} + \beta,
\label{eq:batch_norm}
\end{equation}

where \( \gamma \) and \( \beta \) are learnable parameters \cite{b4}.

Categorical cross-entropy loss is the rule-of-thumb for CNN classification and is formulated as follows with the predicted probabilities \( \hat{y}_i \) and the associated true labels \( y_i \):

\begin{equation}
\text{Loss} = - \sum_{i=1}^{C} y_i \log(\hat{y}_i)
\label{eq:cross_entropy}
\end{equation}

where \( C \) denotes the number of classes. This loss function contributes to effective feature learning that can be applied to a few tasks, e.g., automatic speech recognition (ASR) \cite{b1}.

The usefulness of CNNs as an engine for extracting hierarchical patterns from spectrograms has long been shown. Sainath and Parada \cite{b23} applied CNNs for noisy large vocabulary continuous speech recognition (LVCSR) and showed the model's robust nature, while Abdel-Hamid et al. \cite{b22} demonstrated CNNs' superiority in phoneme recognition. CNNs are useful for speaker ID and verification because CNNs are able to learn speaker-specific features \cite{b15}.

CNNs are tractable for high-bandwidth communication systems since, from the statistical signal processing viewpoint, CNNs are naturally expressive for feature extraction from the perfect specification. Specifically, the signal processing efficiency was highlighted in the future 5G communication system \cite{b3,b7}.

\subsection{Conformers}

To achieve state-of-the-art results for Automatic Speech Recognition (ASR), Gulati et al. \cite{b1} proposed the Conformer architecture, which incorporates convolutional layers within the Transformer design to offer both local and global information concurrently. Transformers are not adaptive to local patterns—something CNNs are apt at capturing—even when they leverage self-attention to model long-range dependencies. The Conformer makes use of these inherent strengths by adding convolutional layers to Transformer blocks.

\begin{figure}[htbp]
\centering
\includegraphics[width=1\linewidth]{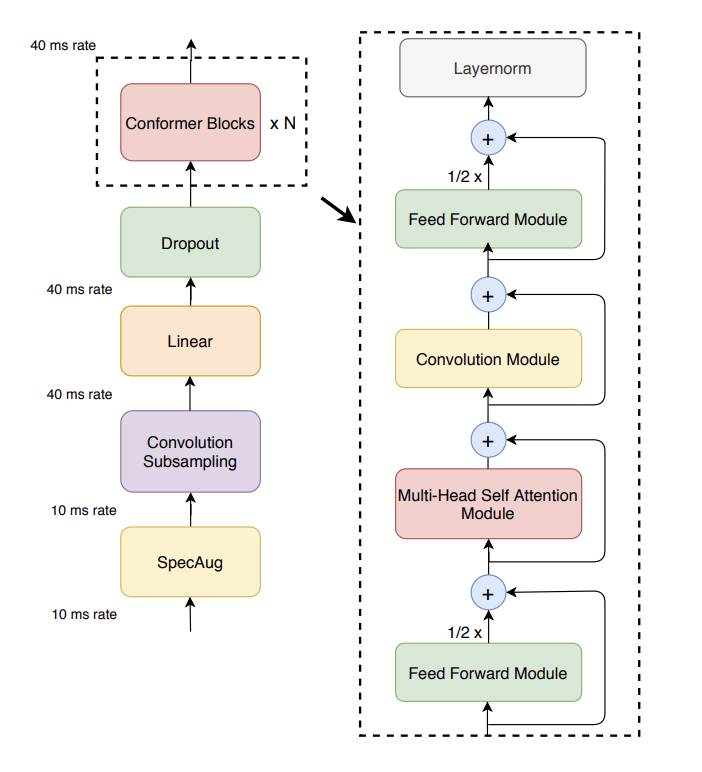}
\caption{Conformer model architecture}
\label{fig:demonstration_selection}
\end{figure}

Each Conformer block's forward pass is:

\begin{align}
\tilde{x}_i &= x_i + \frac{1}{2} \text{FFN}(x_i), \\
x'_i &= \tilde{x}_i + \text{MHSA}(\tilde{x}_i), \\
x''_i &= x'_i + \text{Conv}(x'_i), \\
y_i &= \text{Layernorm}\left(x''_i + \frac{1}{2} \text{FFN}(x''_i)\right)
\label{eq:conformer_block}
\end{align}

where \( x_i \) is the input to the \( i \)-th block. From the above setup, MHSA can learn long-range dependencies as well as short-range dependencies using the convolutional layers to refine local features \cite{b1}.

In Conformer, the convolutional module extracts local dependencies that are necessary for audio processing. For an input sequence \( s(t) \) and a filter \( W \), the convolution operation is:

\begin{equation}
y(t) = \sum_{k=-K}^{K} W(k) \cdot s(t - k)
\label{eq:conv_operation}
\end{equation}

where \( y(t) \) captures local features like phonemes. Conformer employs depthwise separable convolutions to reduce parameter complexity. Depthwise convolution operates per channel:

\begin{equation}
\text{DepthwiseConv}(x) = x * W_d,
\label{eq:depthwise_conv}
\end{equation}

followed by pointwise convolution to merge channels:

\begin{equation}
\text{PointwiseConv}(x) = W_p \cdot x
\label{eq:pointwise_conv}
\end{equation}

MHSA handles global dependencies and performs scaled dot-product attention. For query \( Q \), key \( K \), and value \( V \), attention \( A \) is computed as:

\begin{equation}
A = \text{softmax}\left(\frac{QK^T}{\sqrt{d_k}}\right) V,
\label{eq:attention}
\end{equation}

where \( d_k \) is the dimension of the keys, allowing the model to focus on different parts of the input sequence.

The FFN applies position-wise non-linear transformations, enhancing representational power. Given input \( x \), the FFN is:

\begin{equation}
\text{FFN}(x) = \sigma(W_2 \cdot \text{ReLU}(W_1 \cdot x + b_1) + b_2)
\label{eq:ffn}
\end{equation}

where \( W_1 \) and \( W_2 \) are weights, \( b_1 \) and \( b_2 \) are biases, and \( \sigma \) is an activation function, enabling complex relationship modeling within the sequence.

The Conformer's convolution-augmented Transformer framework is consistent with statistical signal processing concepts, and employs both intrinsic local and extrinsic global statistical characteristics for efficient modelling of the speech signal. This balance has also resulted in better ASR performance due to lower Word Error Rates (WER) in the benchmarks \cite{b1, b5}.

\subsection{Residual Networks (ResNet)}

Residual Networks (ResNet), introduced by He et al. \cite{b25}, address the vanishing gradient issue by introducing residual connections, where some information can go around layers. This skip connection allows stable learning in deep architectures and, hence, ResNet is powerful for complex hierarchical feature extraction tasks, e.g., speech, image processing \cite{b26}.

\begin{figure}[htbp]
\centering
\includegraphics[width=1\linewidth]{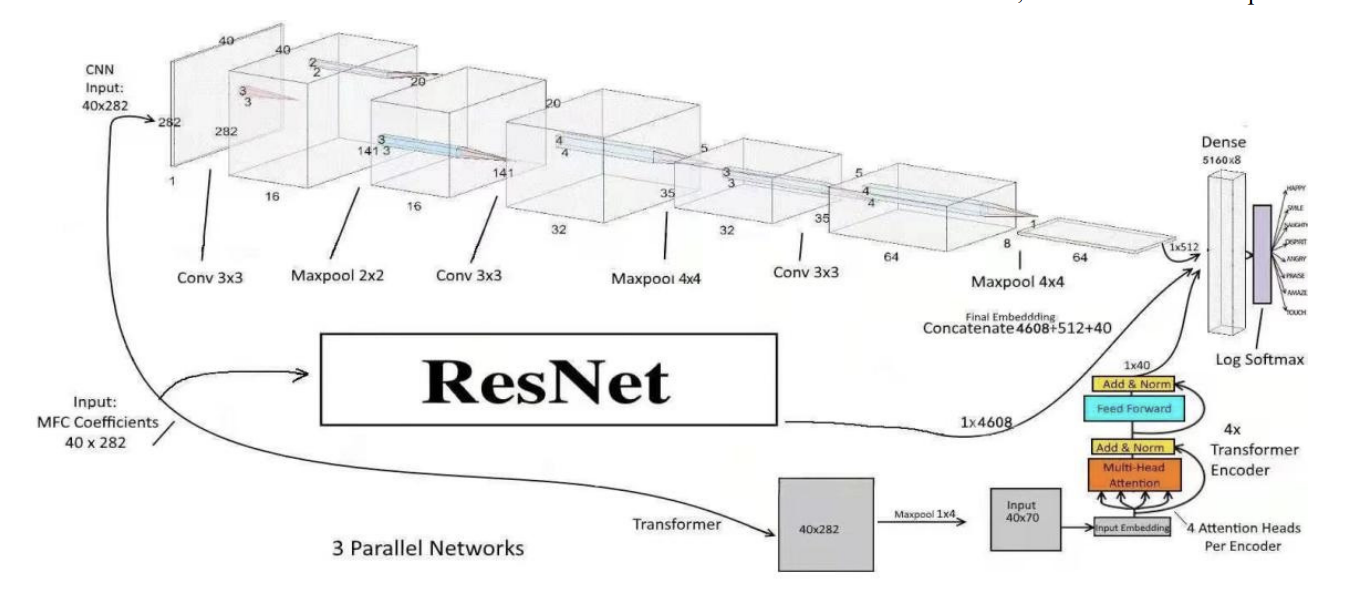}
\caption{ResNet architecture proposed in \cite{b31}}
\label{fig:demonstration_selection}
\end{figure}

In each residual block, the model learns a residual mapping instead of a direct transformation. Given an input \( x \), the output of a residual block is:

\begin{equation}
y = f(x, \{W_i\}) + x
\label{eq:resnet_block}
\end{equation}

where \( f(x, \{W_i\}) \) represents the operations within the residual block, typically consisting of two convolutional layers followed by batch normalization:

\begin{equation}
f(x, \{W_i\}) = \text{\textit{BatchNorm}}(\text{\textit{ReLU}}(\text{\textit{BatchNorm}}(W_1 \cdot x))) + W_2 \cdot x.
\label{eq:residual_mapping}
\end{equation}

The skip connection \( x \) helps maintain gradient flow, addressing the vanishing gradient problem \cite{b27}.

ResNet processes 2D spectrograms with rows representing frequency and columns representing time intervals. The convolution operation is defined as:

\begin{equation}
y[i, j] = \sum_{m=0}^{M-1} \sum_{n=0}^{N-1} W[m, n] \cdot S[i - m, j - n]
\label{eq:conv_operation}
\end{equation}

where \( S[i, j] \) is the spectrogram input at time \( i \) and frequency \( j \), and \( W \) is the convolution kernel. This operation leverages local dependencies necessary for decoding serial audio data.

The residual connection across each block enables the model to learn "innovations" or new information \( f(S, \{W_i\}) \) while maintaining the original signal \( S \):

\begin{equation}
y = f(S, \{W_i\}) + S,
\label{eq:residual_connection}
\end{equation}

where \( S \) denotes the input spectrogram, and \( f(S, \{W_i\}) \) is the transformation within the block. This approach allows the model to capture both temporal and spectral correlations, improving its ability to map phoneme to prosodic information in applications like automatic speech recognition (ASR) and speaker identification \cite{b29}.

The gradient of a layer \( L \) with output \( y_L \) and loss \( \mathcal{L} \) is preserved through the skip connection:

\begin{equation}
\frac{\partial \mathcal{L}}{\partial y_L} = \frac{\partial \mathcal{L}}{\partial y_{L+1}} \cdot \left(1 + \frac{\partial f(y_L)}{\partial y_L}\right).
\label{eq:gradient_flow}
\end{equation}

where \( f(y_L) \) is the residual mapping. This gradient stability allows deep ResNet architectures to learn discriminative, hierarchical features in sequential data such as audio \cite{b28, b30}.

\subsection{Convolutional Recurrent Neural Networks (CRNNs)}

Convolutional Recurrent Neural Networks (CRNNs) integrate Convolutional Neural Networks (CNNs) and Recurrent Neural Networks (RNNs), in particular Long Short-Term Memory (LSTM) units, to improve upon the use of both spatial and temporal dependencies in sequential data, e.g., audio spectrograms \cite{b11}. This architecture is particularly appropriate for problems that need to extract local features as well as long-range temporal dependencies.

Initially, CRNNs apply a series of convolutional layers to extract spatial features from an input spectrogram \( S(i, j) \), where \( i \) and \( j \) denote the time and frequency dimensions. For each convolutional filter \( W(m, n) \), the subsequent feature map \( y(i, j) \) is:

\begin{equation}
y(i, j) = \sum_{m=-M}^{M} \sum_{n=-N}^{N} W(m, n) \cdot S(i - m, j - n)
\label{eq:crnn_conv}
\end{equation}

where \( M \) and \( N \) define the kernel size. Pooling layers, e.g., max pooling, are commonly used to downsample the spatial resolution, preserving the strongest features and minimizing computational complexity:

\begin{equation}
Z(i, j) = \max_{0 \leq m < p} \max_{0 \leq n < q} y(i + m, j + n)
\label{eq:crnn_pooling}
\end{equation}

where \( p \times q \) is the pooling window size.

Following the convolutional feature extraction, the output is reshaped into a sequence structure and fed into an LSTM layer that examines the temporal dependencies of the extracted features. For an input feature sequence \( x_t \) at time step \( t \), the LSTM layer updates its cell state \( c_t \) and hidden state \( h_t \) using the following equations:

\begin{align}
f_t &= \sigma(W_f \cdot h_{t-1} + U_f \cdot x_t + b_f) \\
i_t &= \sigma(W_i \cdot h_{t-1} + U_i \cdot x_t + b_i) \\
\tilde{c}_t &= \tanh(W_c \cdot h_{t-1} + U_c \cdot x_t + b_c) \\
c_t &= f_t \cdot c_{t-1} + i_t \cdot \tilde{c}_t \\
h_t &= o_t \cdot \tanh(c_t)
\label{eq:lstm_equations}
\end{align}

where \( f_t \), \( i_t \), and \( o_t \) are the forget, input, and output gates respectively; \( W \) and \( U \) are weight matrices, and \( b \) is the bias vector. This transformation sequence allows the LSTM to learn long-term dependencies present in the input data \cite{b33}.

The final output from the LSTM layer, \( h_t \), represents the processed temporal features and is passed to a fully connected layer with softmax activation for classification. The softmax function (class probability prediction) is expressed as:

\begin{equation}
\hat{y}_i = \frac{e^{z_i}}{\sum_{j=1}^{C} e^{z_j}},
\label{eq:softmax}
\end{equation}

where \( \hat{y}_i \) is the probability of class \( i \), \( z_i \) is the logit for class \( i \), and \( C \) is the total number of classes.

CRNNs are optimized for speech processing tasks (e.g., automatic speech recognition and speaker identification) by successfully capturing spatial and temporal dependencies in the audio data \cite{b13, b18}. Convolutional layers learn local time-frequency representations, while LSTM layers preserve sequential dependencies, making CRNNs effective for challenging audio and sequential data representations.

\section{Applications}

\subsection{Speech Recognition}

Speech recognition is the task of translating spoken language into text or machine understandable commands. It is another fundamental one in the area of speech signal processing, and has been extensively used in virtual assistant, transcription, and voice-activated systems, etc. The state-of-the-art performance of speech recognition systems is significantly enhanced by convolutional-based system architectures through effectively capturing local and global aspects of the speech signal.

\begin{figure}[!t]
    \centering
    \includegraphics[width=0.45\textwidth]{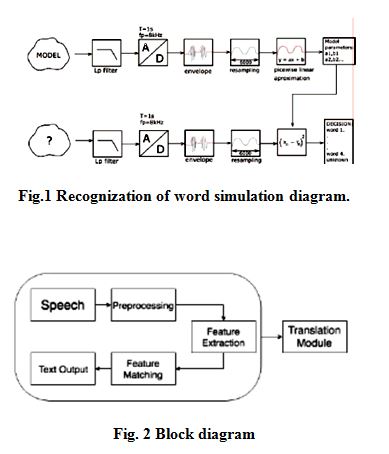}
    \caption{Speech Signal Processing Pipeline}
    \label{fig:speech_processing}
\end{figure}

The overall objective in speech recognition is to represent the probability of a word sequence $W = \{w_1, w_2, \dots, w_T\}$ conditioned on a set of acoustic features $X = \{x_1, x_2, \dots, x_T\}$. This is often formulated using Bayesian decision theory:

\begin{equation}
    P(W|X) = \frac{P(X|W)P(W)}{P(X)}
    \label{eq:bayes}
\end{equation}

And where $P(X|W)$, $P(W)$, and $P(X)$ are the acoustic and language models and evidence with the possibility to be set to zero during decoding.

Convolutional Neural Networks (CNNs) have been used before to model $P(X|W)$ learning hierarchical representations of input features \cite{b19}. Convolutional neural networks (CNNs) extract local temporal and spectral correlations in speech signals by means of convolutional filters. For speech recognition Convolutional Neural Networks (CNNs), the convolution process is represented as:

\begin{equation}
    y_{i,j} = \sigma\left(\sum_{m=-M}^{M} \sum_{n=-N}^{N} w_{m,n} \cdot x_{i+m,j+n} + b\right)
    \label{eq:cnn_convolution}
\end{equation}

where $y_{i,j}$ is the output feature map, $x_{i,j}$ is the input feature map, $w_{m,n}$ are the weights of the convolutional kernel, $b$ is the bias term, and $\sigma$ is the activation function.

Conformers, introduced by Gulati et al. \cite{b1} refine CNNs by incorporating self-attention mechanisms that can learn long-range information without the loss of the ability to model local features in the convolution. The Conformer block embeds convolutional blocks and Transformer layers, and the output can be expressed as:

\begin{equation}
    \begin{aligned}
    \text{\textit{ConformerBlock}}(X) = &\text{\textit{F}}_{\frac{1}{2}}(X) + \text{\textit{M}}(X) + \\
    &\text{\textit{C}}(X) + \text{\textit{F}}_{\frac{1}{2}}(X)
    \end{aligned}
    \label{eq:conformer_block}
\end{equation}

where $\text{F}_{\text{half}}$ is a feed-forward network with half the step size, $\text{M}$ is multi-head self-attention, and $\text{C}$ represents the convolution module.

The depthwise separable convolutions unit in the architecture of the Conformer learns relationships between neighbors:

\begin{equation}
    Y = X + \text{\textit{PointwiseConv}}(\text{\textit{GLU}}(\text{\textit{DepthwiseConv}}(X)))
    \label{eq:conformer_conv}
\end{equation}

Where the convolution filter is different for each input channel in $\text{DepthwiseConv}$, the gated linear unit activation $\text{GLU}$, and combined channels in $\text{PointwiseConv}$.

Convolutional Recurrent Neural Networks (CRNNs) are a mixture of CNNs and recurrent units (e.g., Long Short-Term Memory (LSTM) networks) to extract spatial and temporal features in speech signals (see \cite{b11}). Local features are extracted by the convolutional neural network layers and long-term temporal context is learned by the recurrent layers. The CRNN output can be described as:

\begin{equation}
    H_t = \text{LSTM}(\text{CNN}(X_t), H_{t-1})
    \label{eq:crnn}
\end{equation}

Where $X_t$ is the input at time $t$, $H_t$ is the hidden state, and $\text{CNN}(X_t)$ outputs the feature representation of input.

Residual Networks (ResNets) allow the training of extremely deep CNNs by adding residual connections to overcome the vanishing gradient problem \cite{b7}. The residual connection is derived by summing over the input of a layer to the output of a layer:

\begin{equation}
    Y = F(X, W) + X
    \label{eq:resnet}
\end{equation}

Where $F(X, W)$ is the output of the convolutional layers, with weight $W$, and $X$ is the input to the residual block.

In the area of statistical signal processing, these architectures help to estimate the acoustic model $P(X|W)$ more accurately by means of robust feature extraction and modeling. They enhance the discriminative capability of speech recognition systems (e.g., in the presence of noise or limited training data) \cite{b5}, \cite{b19}.

For example, Alami et al. \cite{b19} showed that by learning invariant features in spectrograms, CNNs can perform noise-robust speech recognition. Gulati et al. \cite{b1} demonstrated that Conformers outperformed conventional CNNs and Transformers by leveraging both convolution and self-attention capabilities.

In addition, the combination of statistical approaches and deep learning architectures has resulted in hybrid models that continue to enhance performance. For example, hybrid Hidden Markov Model (HMM)-DNN architectures employ convolutional structures to approximate emission probabilities of HMMs \cite{b20}.

\subsection{Speaker Identification}

Speaker identification is the process of identifying the identity of a speaker from vocal characteristics. In applications ranging from security systems, personalized user interfaces, and forensic investigation, this task is extremely crucial. Convolution-based architectures have been instrumental in enhancing the accuracies and robustness of speech-recognition systems by capturing the unique variability in each person's voice.

\begin{figure}[!t]
    \centering
    \includegraphics[width=0.4\textwidth]{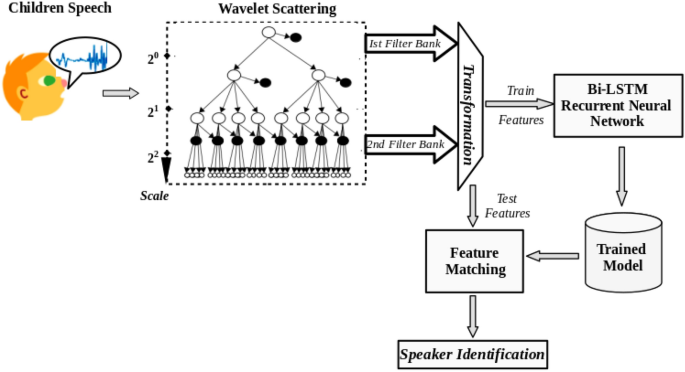}
    \caption{Speaker Identification Pipeline}
    \label{fig:speaker_identification}
\end{figure}

The main task in speaker identification is to estimate the probability $P(S|X)$ that a certain speaker $S$ is spoken for a set of acoustic feature sequences $X = \{x_1, x_2, \dots, x_t\}$. This can be formulated using Bayesian inference as:

\begin{equation}
    P(S|X) = \frac{P(X|S)P(S)}{P(X)}
    \label{eq:speaker_bayes}
\end{equation}

Here, $P(X|S)$, $P(S)$, and $P(X)$ are the acoustic model, prior probability of the speaker, and evidence, respectively.

Convolutional Neural Networks (CNNs) have been applied with success to model $P(X|S)$ through learning hierarchical features of the input acoustic representations, such as Mel-Frequency Cepstral Coefficients (MFCCs) \cite{b15}. The convolutional operation of CNNs for use in speaker identification can be written as:

\begin{equation}
    Y_{i,j} = \sigma\left(\sum_{m=-M}^{M} \sum_{n=-N}^{N} w_{m,n} \cdot x_{i+m, j+n} + b\right)
    \label{eq:cnn_speaker}
\end{equation}

Where $Y_{i,j}$ is the output feature map $x_{i,j}$ is the input feature map, $w_{m,n}$ is the convolutional kernel weights, $b$ is the bias, and $\sigma$ is the activation function.

Convolutional Recurrent Neural Networks (CRNNs) approximate spatial and temporal relationships in speech signals by combining convolutional neural networks (CNNs) and recurrent networks (e.g., Long Short-Term Memory (LSTM) networks) \cite{b11}. The architecture of the CRNN model analyzes input sequence $X_t$ at time $t$ as:

\begin{equation}
    H_t = \text{LSTM}(\text{CNN}(X_t), H_{t-1})
    \label{eq:crnn_speaker}
\end{equation}

Specifically, in which $H_t$ (hidden state) and $\text{CNN}(X_t)$ (extracting spatial information using the convolutional neural network) are applied.

Residual networks (ResNets) train deep CNNs by adding "residual connections" that solve the vanishing gradient problem and enable training of models with increased depth without performance loss \cite{b20}. The residual connection is mathematically represented as:

\begin{equation}
    Y = F(X, W) + X
    \label{eq:resnet_speaker}
\end{equation}

Where $F(X, W)$ is the output of convolutional layers with weights $W$, where $X$ is the input of residual block.

Gaussian Mixture Models (GMMs) are frequently employed together with CNNs for estimating the distribution of acoustic features of each speaker \cite{b17}. The probability $P(X|S)$ can be expressed as:

\begin{equation}
    P(X|S) = \sum_{k=1}^{K} \omega_k \mathcal{N}(X|\mu_k, \Sigma_k)
    \label{eq:gmm}
\end{equation}

Here, $\omega_k$ are the mixing weights, and $\mathcal{N}(X|\mu_k, \Sigma_k)$ are the components of the Gaussian distributions with mean $\mu_k$ and covariance $\Sigma_k$.

In noisy conditions, convolutional-based models augmented with statistical signal processing (SSP) techniques have been demonstrated to be highly robust. For instance, Na et al. As demonstrated in \cite{b15}, CNNs with noise-resistant feature extraction techniques dramatically improve the effectiveness of speaker recognition in the presence of acoustic background noise. In addition, by combining GMM and CNN, hybrid models have enabled real-time speaker recognition at good accuracy \cite{b17}.

Generally, convolution-based architectures are naturally suited to statistical signal processing paradigms for the ability to learn robust features as well as model subject speaker properties.

\subsection{Emotion Detection}

Speech emotion recognition is the task by which the emotional state of a person is decoded from vocalization. The role of the task is significant for applications such as human-computer interaction, mental health monitoring and automatic customer service. Conv-based architectures have now allowed impressive performance gains for emotion detection systems regarding both accuracy and robustness, since they are capable of modeling the patterns and variability of speech signals underlying a wide range of emotions.

The objective in emotion detection is to model the probability $P(E|X)$ of an emotion $E$ given an acoustic feature sequence $X = \{x_1, x_2, \dots, x_T\}$. This can be formulated using Bayesian inference as:

\begin{equation}
    P(E|X) = \frac{P(X|E)P(E)}{P(X)}
    \label{eq:emotion_bayes}
\end{equation}

Where $P(X|E)$ denotes the emotional acoustic model, $P(E)$ the prior probability of the emotion and $P(X)$ the evidence.

Convolutional Neural Networks (CNNs) have been widely utilized to encode $P(X|E)$ learning hierarchical features from input acoustic representations such as Mel-Frequency Cepstral Coefficients (MFCCs) \cite{b13}. The convolution function in CNNs (for emotion detection) can be represented as:

\begin{equation}
    y_{i,j} = \sigma\left(\sum_{m=-M}^{M} \sum_{n=-N}^{N} w_{m,n} \cdot x_{i+m, j+n} + b\right)
    \label{eq:cnn_emotion}
\end{equation}

Where $y_{i,j}$ is the output feature map, $x_{i,j}$ is the input feature map, $w_{m,n}$ are the weights of convolutional kernels, $b$ is the bias, and $\sigma$ is the activation function.

CNN-LSTM networks use convolution units with Long Short Term Memory (LSTM) units for joint acquisition of spatial and temporal features in speech signals \cite{b16}. The hybrid architecture views the input sequence $X_t$ at time $t$ as:

\begin{equation}
    H_t = \text{LSTM}(\text{CNN}(X_t), H_{t-1})
    \label{eq:cnn_lstm_emotion}
\end{equation}

where $H_t$ is the hidden state at time $t$, and the spatial features are obtained from the input with $\text{CNN}(X_t)$.

Residual networks (ResNets) enhance the performances of deep Convolutional Neural Networks (CNNs) just by inserting residual connections, enabling deep network training without performance degrading \cite{b7}. Residual connection of emotion detection can be described as:

\begin{equation}
    Y = F(X, W) + X
    \label{eq:resnet_emotion}
\end{equation}

In which $F(X, W)$ is the output obtained from convolutional layers with weights $W$ and $X$ is the input provided to the residual block.

Activation functions and normalization techniques in emotion detection based deep learning models are commonly applied to ensure training stability and convergence. E.g., information flow on the network is restricted using the Gated Linear Unit (GLU) activation function:

\begin{equation}
    \text{GLU}(a, b) = a \otimes \sigma(b)
    \label{eq:glu}
\end{equation}

In which $a$ and $b$ are input tensors, $\otimes$ denotes element-wise multiplication, and $\sigma$ is the sigmoid function.

In the area of statistical signal processing, such convolution-based architectures enhance the process of feature extraction through considering the distribution of so-called emotional features in the speech signal. Wang et al. \cite{b18} survey various deep learning approaches for emotion recognition, highlighting the effectiveness of CNNs in capturing discriminative features. Prabhu and Raj\textsuperscript{1} demonstrated that CNNs can be applied to discriminative fine emotional cues by the learning of abstract features from the convolution of the spectrogram.

Furthermore, CNN-LSTM networks as proposed by Kumar and Sharma \cite{b16} incorporate temporal dynamics of speech for an improvement of the emotion classification performance. These hybrid models combine the local feature extraction capabilities of CNNs with the sequence modeling strengths of LSTMs, providing a comprehensive framework for emotion detection.

Conceptually, convolution-based architectures are consistent with principles of statistical signal processing by providing discriminative feature learning and good models of emotionality for speech signals. These developments have resulted in more precise and trustworthy emotion detection systems capable of performing robustly in heterogeneous and dynamic environments.

\section{Comparative Analysis of the Architectures}

Model size, accuracy, speed, and training cost may all be used to compare the four models. The findings from the \textit{VoxForge} and \textit{Voxlingua6} datasets \cite{b9} served as the basis for the related analysis. \textit{VoxForge} includes speech data in English, German, Russian, Italian, Spanish, and French, whereas \textit{Voxlingua6} adds more speaker and language variability. Both datasets include speech data from various languages. Data statistics for each dataset are shown in Table \ref{tab:datasetstransposed}.

\begin{table}[htbp]
\caption{Data Statistics on \textit{VoxForge} and \textit{Voxlingua6} Datasets}
\begin{center}
\begin{tabular}{|c|c|c|c|}
\hline
\textbf{Characteristic} & \textbf{Train} & \textbf{Validation} & \textbf{Test} \\
\hline
English & 291 spk & 27 spk & 41 spk \\
\hline
German & 90 spk & 7 spk & 7 spk \\
\hline
Russian & 193 spk & 7 spk & 8 spk \\
\hline
Italian & 152 spk & 16 spk & 15 spk \\
\hline
Spanish & 280 spk & 10 spk & 18 spk \\
\hline
French & 195 spk & 8 spk & 9 spk \\
\hline
\end{tabular}
\label{tab:datasetstransposed}
\end{center}
\end{table}

\subsection{Training Cost}

The amount of processing power required to train each model is known as the training cost. Models with more parameters often demand more time and processing power. The CNN architecture is light (6 million parameters) and hence versatile, whereas the Conformer architecture is heavy (15.5 million parameters), as seen in Table \ref{tab:modelsizes} from \cite{b9}. Convolutional and recurrent layers are combined in CRNN, which contains over 19.5 million parameters. Because of the intricate convolutional self-attention mechanism, the Conformer and CRNN require additional processing resources, particularly in noisy and multilingual environments like \textit{Voxlingua6}. The memory space and computational cost during deployment are directly impacted by the model size, or the number of parameters.

\begin{table}[htbp]
\caption{Total Number of Parameters in Each Model}
\begin{center}
\begin{tabular}{|c|c|}
\hline
\textbf{Model} & \textbf{\# of Parameters (million)} \\
\hline
CNN & 6.0 \\
\hline
CRNN & 19.5 \\
\hline
Residual Network & 23.5 \\
\hline
Conformer & 15.5 \\
\hline
\end{tabular}
\label{tab:modelsizes}
\end{center}
\end{table}

\begin{figure}[htbp]
\centering
\includegraphics[width=0.8\linewidth]{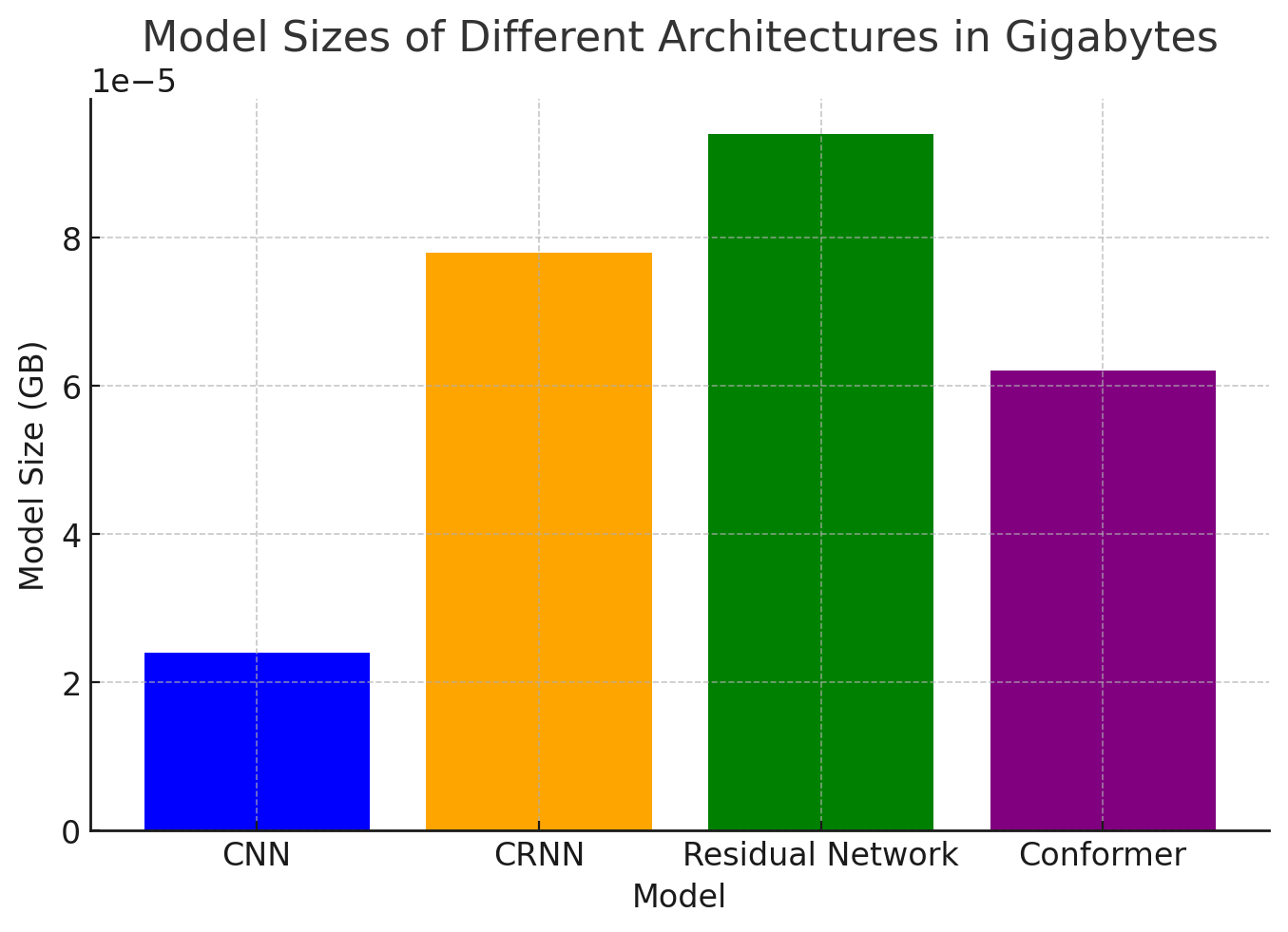}
\caption{Model Parameter Sizes}
\label{fig:demonstrationselection}
\end{figure}

\subsection{Accuracy}

The Conformer on average outperforms the other architectures (error rate 5.27\% on the \textit{Voxlingua6} Dev set). CNNs follow closely with an error rate of 7.18\%, while the CRNN and Residual Network perform slightly worse, reflecting the trade-off between model complexity and accuracy.

\begin{table}[htbp]
\caption{Error Rates [\%] of Models on \textit{Voxlingua6} Dev Set}
\begin{center}
\begin{tabular}{|c|c|}
\hline
\textbf{Model} & \textbf{Error Rate [\%]} \\
\hline
CNN & 7.18 \\
\hline
CRNN & 11.35 \\
\hline
Residual Network & 8.56 \\
\hline
Conformer & 5.27 \\
\hline
\end{tabular}
\label{tab:errorrates}
\end{center}
\end{table}

\begin{figure}[htbp]
\centering
\includegraphics[width=0.8\linewidth]{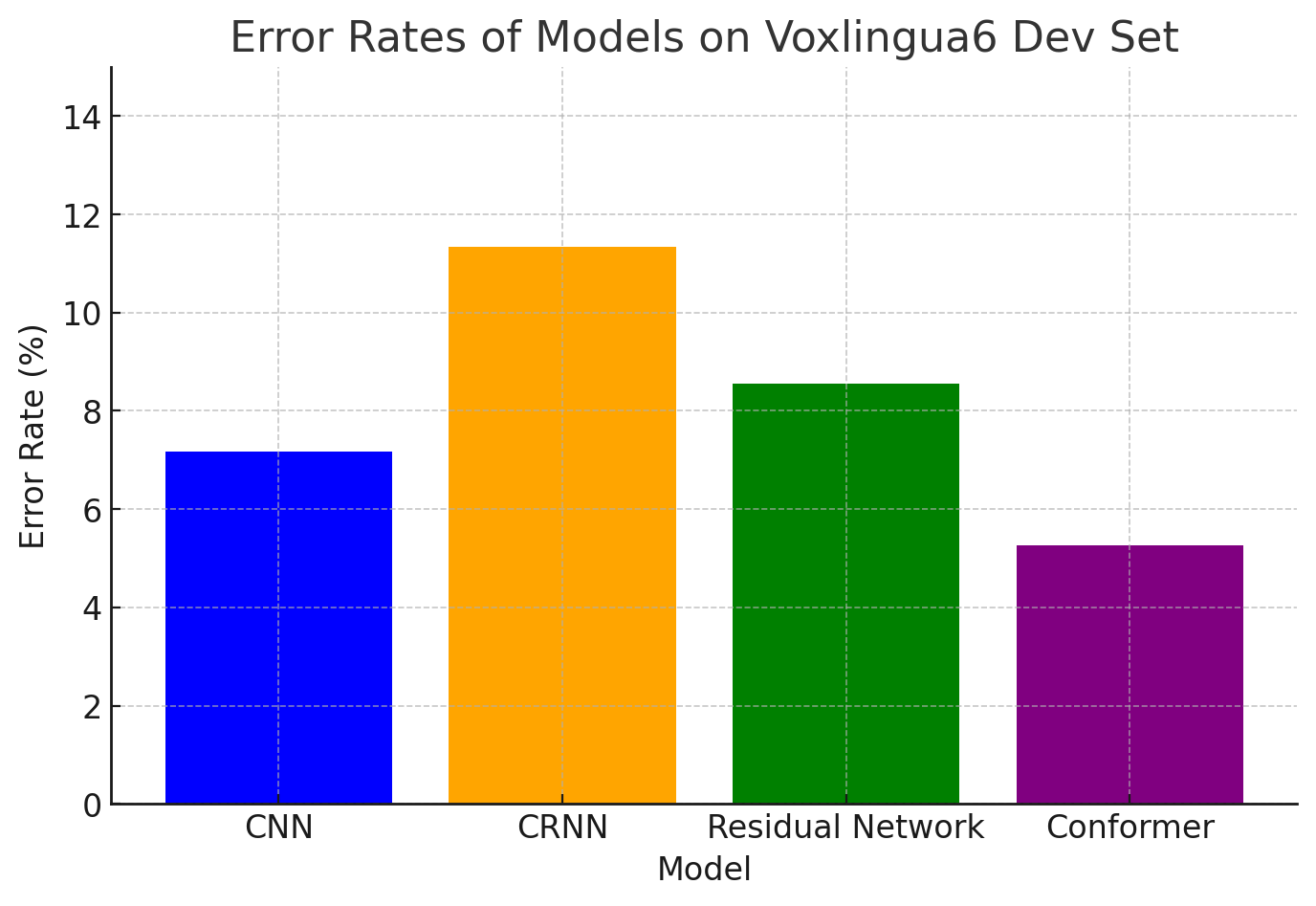}
\caption{Error Rates Comparison}
\label{fig:demonstration_selection}
\end{figure}

\subsection{Speed}

In terms of speed, the CNN is most efficient for real-time applications, while the Conformer strikes a balance between speed and accuracy, making it suitable for slightly delayed yet accurate systems.

\section{Conclusion}

In order to accomplish speech signal processing tasks including speaker identification, emotion detection, and voice recognition, this article compared CNN, Conformer, CRNN, and Residual Network designs. We tested these models on training cost, model size, accuracy, and inference speed using the \textit{VoxForge} and \textit{Voxlingua6} datasets. We discovered that Conformers are more accurate than CNNs, which are more suited for low-resource situations because of their lower size \cite{b1}, \cite{b9}.

Future studies will concentrate on improving noise resistance, developing effective, low-latency models for real-time applications, and reducing architectural complexity to make them easier to utilise on devices with constrained resources. Investigating hybrid configurations that combine convolution and self-supervised learning \cite{b4} may result in additional advancements by striking a balance between model complexity and performance.

\vspace{12pt}

\end{document}